\documentclass[apj]{emulateapj}

\def\kms{{\rm\,km\,s^{-1}}}
\def\kpc{{\rm\,kpc}}
\def\deg{^\circ}

\shorttitle{The enigmatic pair of dSphs Leo IV and Leo V}
\shortauthors{de Jong et al.}

\begin{document}

\title{The enigmatic pair of dwarf galaxies Leo IV
  and Leo V: coincidence or common origin?}

\author{Jelte T. A. de Jong\altaffilmark{1},
 Nicolas F. Martin\altaffilmark{1}, Hans-Walter Rix\altaffilmark{1},
 Kester W. Smith\altaffilmark{1}, Shoko Jin\altaffilmark{2,3}, Andrea
 V. Macci\`o\altaffilmark{1}}

\email{dejong@mpia.de}

\altaffiltext{1}{Max-Planck-Institut f\"{u}r Astronomie,
 K\"{o}nigstuhl 17, D-69117 Heidelberg, Germany}
\altaffiltext{2}{Astronomisches Rechen-Institut, Zentrum f\"ur
 Astronomie der Universit\"at Heidelberg, M\"onchhofstra\ss e 12-14,
 D-69120 Heidelberg, Germany}
\altaffiltext{3}{Alexander von Humboldt research fellow}

\begin{abstract}
We have obtained deep photometry in two $1\deg\times1\deg$ fields
covering the close pair of dwarf spheroidal galaxies Leo IV and Leo V
and part of the area in between. From the distribution of likely red
giant branch and horizontal branch stars in the data set, we find that
both Leo IV and Leo V are significantly larger than indicated by
previous measurements based on shallower data. With a half-light
radius of $r_h$=4\farcm6$\pm0\farcm8$ (206$\pm$36 pc) and
$r_h$=2\farcm6$\pm$0\farcm6 (133$\pm$31 pc), respectively, both
systems are now well within the physical size bracket of typical dwarf
spheroidal Milky Way satellites. Both are also found to be
significantly elongated with an ellipticity of $\epsilon\simeq0.5$, a
characteristic shared by many of the fainter ($M_V>-8$) Milky Way
dwarf spheroidals.  The large spatial extent of our survey allows
us to search for extra-tidal features in the area between the two
dwarf galaxies with unprecedented sensitivity.  The spatial
distribution of candidate red giant branch and horizontal branch stars
in this region is found to be non-uniform at the $\sim3\sigma$
level. Interestingly, this substructure is aligned along the direction
connecting the two systems, indicative of a possible `bridge' of
extra-tidal material.  Fitting the stellar distribution with a linear
Gaussian model yields a significance of 4$\sigma$ for this
overdensity, a most likely FWHM of $\sim$16 arcmin and a central
surface brightness of $\simeq$32 mag arcsec$^{-2}$.  We investigate
different scenarios to explain the close proximity of Leo IV and Leo V
and the possible tidal bridge between them. Orbit calculations
demonstrate that the two systems cannot share the exact same orbit,
while a compromise orbit does not approach the Galactic center more
than $\sim$160 kpc, rendering it unlikely that they are remnants of a
single disrupted progenitor. A comparison with cosmological
simulations shows that a chance collision between unrelated subhalos
is negligibly small. Given their relative distance and velocity, Leo
IV and Leo V could be a bound `tumbling pair', if their combined mass
exceeds 8$\pm4 \times 10^9$ M$_\sun$. The scenario of an internally
interacting pair that fell into the Milky Way together appears to be
the most viable explanation for this close celestial companionship.
\end{abstract}


\section{Introduction}
\label{sec:intro}

In the past four years, the Sloan Digital Sky Survey (SDSS) has
revealed satellite galaxies of the Milky Way (MW) that are up to 100
times fainter than those known before, more than doubling the number
of known satellites. Most of these new discoveries are dwarf
spheroidal galaxies (dSph) that are dark-matter dominated objects
\citep{spectro1,spectro2}, yet have surprisingly complex star
formation histories \citep{sdssmatch} and/or morphologies
\citep[e.g.][]{lbtherc,lbtcvnI,lbtleoT}.  Thanks to these new objects
and the uniform sky coverage provided by SDSS, it is now possible to
probe the faint end of the satellite galaxy luminosity function
\citep[e.g.][]{koposov08}. This extended data set has enabled a
detailed comparison with galaxy formation models, leading to better
agreement between models and observation, thereby largely solving the
`missing satellite problem', one of the stumbling blocks for current
cosmological models on small scales
\citep[e.g.][]{koposov09,maccio09}.  Recent studies of the structural
properties of these newly discovered dwarf galaxies based on SDSS data
have revealed that they are, as a class, more elongated than their
brighter counterparts \citep{MLfits}, indicating that some systems may
be tidally disrupted by the MW.  It is as yet unclear whether these
objects represent a lower mass limit to the galaxy formation process,
or whether their stellar bodies are remnants of originally larger
systems.

In this context, the recently discovered systems Leo IV \citep{5pack}
and Leo V \citep{leoV} are particularly interesting, as they are
separated by less than $3.0\deg$ on the sky with fairly similar
distances ($154\pm5\kpc$ and $\sim180\kpc$, respectively;
\citealt{moretti09,leoV}) and radial velocities ($10.1\pm1.4\kms$ and
$60.8\pm3.1\kms$, respectively, with respect to the Galactic Standard
of Rest; \citealt{spectro2,leoV}), implying that they might be related
to a single disrupting or disrupted progenitor. \cite{leoV} also found
the distribution of blue horizontal branch (BHB) stars in Leo V to be
elongated and even detected two candidate red giant branch (RGB) stars
at $\sim$13\arcmin~North of the system, well beyond the estimated
tidal radius. Futhermore, with its estimated half-light radius of
$\sim$40 pc \citep{leoV}, Leo V appears to fall in the `size gap'
between globular clusters and classical dwarf spheroidal galaxies
\cite[cf.][]{gilmore07}.

\begin{figure*}[ht]
\epsscale{1.0}
\plotone{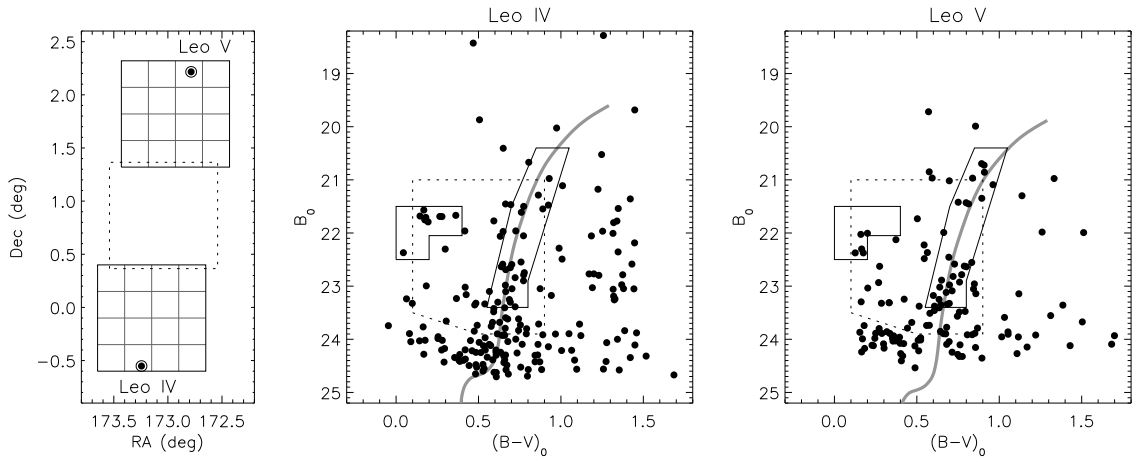}
\caption{ Layout of observed fields and color-magnitude diagrams. {\it
Left:} Black solid lines outline the two $1\deg\times1\deg$ fields
that were obtained, with the gray lines showing the individual
chips. The dashed square outlines the third field that was lost
to bad weather. The locations of Leo IV and Leo V are indicated.  {\it
Middle:} Color-magnitude diagram of Leo IV within 3\arcmin~ from the
center. Overplotted is a 14 Gyr isochrone with [Fe/H]=$-$2.3 and
$m-M$=20.94. Two selection boxes to select HB and RGB stars are shown
with solid outlines, while a third selection box used for the
maximum-likelihood fitting procedure is indicated with a dashed
outline. 
{\it Right:} As the middle panel, but for Leo V. Here the isochrone
has $m-M$=21.23. In both cases the isochrones are from
\cite{dotter08}.  }
\label{fig:data}
\end{figure*}

We have obtained deep photometry in two $1\deg\times1\deg$ fields,
covering both Leo IV and V as well as much of the area in between
them. These new data enable a re-determination of the structural
parameters for both galaxies, as well as a sensitive search for
possible extra-tidal material connecting the two systems. The
remainder of this paper is structured as follows. In \S \ref{sec:data}
the data and data reduction process are detailed. Revised structural
parameters for Leo IV and Leo V are presented in \S
\ref{sec:structpar}. The different ways in which we search for
evidence of extra-tidal stars in between the two systems is described
in \S \ref{sec:stream}, and in \S \ref{sec:discussion} several
scenarios to explain the relation between them are discussed. A short
summary and our conclusions are presented in \S \ref{sec:conclusions}.

\section{Data}
\label{sec:data}

Deep, wide-field photometry in the Johnson $B$ and $V$ filters was
obtained for two $1\deg\times1\deg$ fields, using the LAICA imager on
the 3.5m telescope at the Calar Alto observatory. Four 4k$\times$4k
CCDs with a pixel size of 0\farcs22 are positioned such that a mosaic
of 4 pointings provides a $1\deg\times1\deg$ field of
view. Observations were carried out during the nights of February 24th
and 25th 2009, under photometric conditions with seeing ranging from
1\farcs2 to 1\farcs8.  Total exposure times in $B$ and $V$ were 1950s
and 2600s, divided over single exposures of 650s. The locations of the
fields are shown in the left panel of Figure \ref{fig:data}; a planned
central field could not be obtained due to bad weather.

\begin{figure}[ht]
\epsscale{1.0}
\plotone{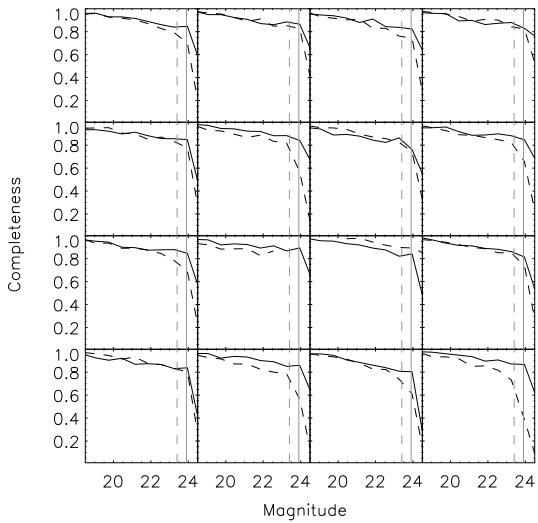}
\caption{Completeness as a function of magnitude for all CCDs in the
  1\degr$\times$1\degr~ field containing Leo IV. The solid line is for
  the $B$-band, and the dashed line for the $V$-band. Gray vertical lines
  indicate the magnitude limits used for the maximum-likelihood
  analysis of the stellar spatial distribution: 23.9 for $B_0$ (solid)
  and 23.4 for $V_0$ (dashed).}
\label{fig:completenessC}
\end{figure}

\begin{figure}[ht]
\epsscale{1.0}
\plotone{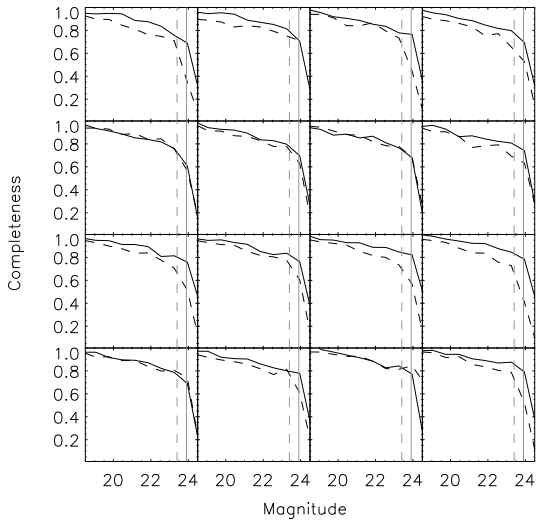}
\caption{ As Fig. \ref{fig:completenessC} but for the field containing
  Leo V. }
\label{fig:completenessA}
\end{figure}

After bias subtraction and flatfielding, the data from the individual
chips were astrometrically corrected and stacked using the {\it SCAMP}
\citep{scamp} and {\it SWARP} software\footnote{{\it SCAMP} and {\it
SWARP} can be obtained from http://terapix.iap.fr}, developed by
Terapix. A chip-by-chip analysis of the data was necessary because of
seeing differences between different pointings and PSF distortions.
Object detection was done using SExtractor \citep{sextractor} and
subsequent PSF-fitting photometry with the {\it daophot} package in
IRAF. By cross-matching objects with SDSS DR7 \citep{dr7} and
converting SDSS magnitudes to $B$ and $V$ using the transformations
from \cite{jester05}, the photometry was calibrated to SDSS. The
photometry of all stars is made available in Table \ref{tab:data},
which contains the coordinates, magnitudes and their uncertainties,
and the extinction at the position of each star interpolated from the
extinction maps of \cite{sfd}.  Using these extinction values, all
stars were individually corrected for the limited foreground
extinction towards these fields ($E(B-V)\sim0.1$). The final
calibrated and dereddened magnitudes will hereafter be referred to as
$B_0$ and $V_0$.  Color-magnitude diagrams (CMD) of Leo IV and V
within radii of 3\arcmin~are presented in Figure
\ref{fig:data}. Artificial star tests to assess the completeness in
all CCDs were done by placing 4\,500 sources spread over a magnitude
range of 17 to 25 in each CCD and for each filter and analysing them
following the exact same procedure. The completenesses derived are
shown in Figure \ref{fig:completenessC} for the Leo IV field and in
Figure \ref{fig:completenessA} for the Leo V field. Seeing variations
cause the completeness to be non-uniform, but $\sim$80\% completeness
is reached to at least $B_0=V_0\simeq$23.5 over the whole field, or
approximately 1.5 magnitudes fainter than the SDSS.

\begin{deluxetable*}{cccccccc}
\tablecaption{Photometric data}
\tablewidth{0pt} 
\tablehead{ \colhead{Chip ID$^*$} & \colhead{$\alpha$} &
  \colhead{$\delta$} & \colhead{$B$} & \colhead{$\sigma_B$} &
  \colhead{$V$} & \colhead{$\sigma_V$} & \colhead{E(B-V)}\\
 ~ & (\degr~J2000.0) & (\degr~J2000.0) & (mag) & (mag) & (mag) & (mag) & (mag)}
\startdata
0\_11 & 173.33533 & 2.05084 & 23.302 & 0.042 & 21.902 & 0.014 & 0.028 \\
0\_11 & 173.23705 & 2.05027 & 23.603 & 0.048 & 22.295 & 0.021 & 0.028 \\
0\_11 & 173.28453 & 2.05040 & 23.531 & 0.042 & 23.185 & 0.047 & 0.028 \\
0\_11 & 173.29111 & 2.04999 & 24.160 & 0.071 & 23.551 & 0.074 & 0.028 \\
0\_11 & 173.27151 & 2.05098 & 23.059 & 0.028 & 22.804 & 0.032 & 0.028 \\
\enddata
\tablecomments{
The complete table is available online.\\
$^*$~The first digit of the chip ID indicates the field,
  where a `0' stands for the Northern field containing Leo V and a `1'
  for the Southern field containing Leo IV; the integers following the
  underscore signify the exposure and the chip number, respectively,
  both running from 1 to 4.}
\label{tab:data}
\end{deluxetable*}

\section{Revised properties of Leo IV and Leo V}
\label{sec:structpar}

An accurate distance to Leo IV has been determined by \cite{moretti09}
using RR Lyrae stars.  Although the absolute calibration of our
photometry would introduce uncertainties in the determination of the
distance to Leo V, the luminosity of its HB stars can be compared to
the luminosity of the HB in Leo IV, providing a distance offset with
respect to the latter.  The stars on the horizontal part of the
horizontal branch (HB) of Leo IV have a mean $B_0=$21.78$\pm$0.06,
while for Leo V we find $B_0=$22.06$\pm$0.04. Assuming a distance
modulus of 20.94$\pm$0.07 \citep[154$\pm$5 kpc,][]{moretti09} for Leo
IV, this gives a distance modulus of 21.22$\pm$0.10, or a distance of
175$\pm$9 kpc, for Leo V.

Based on our new photometry, we re-determine the structural parameters
of the two dwarf galaxies using the algorithm developed and detailed
in \citet{MLfits}. We refer the reader to this paper for details but
recall that this maximum likelihood (ML) algorithm uses the spatial
positions of stars that are plausible members of a dwarf galaxy
(selected by a rough color-magnitude cut) to determine the best
elliptical, exponential surface density profile of the satellite, and
fit for the fore/background stars at the same time. Given the
positions of Leo IV and V near the edge of the observed fields, we use
the adapted version of the algorithm presented in \S3.3 of
\citet{martin09}. This version accounts for holes and gaps in
data sets by iteratively populating un-observed regions with a stellar
density that follows the best model from the previous iteration. The
contribution of this stochastic method to the uncertainties on the
model parameters is determined via a Monte Carlo approach, in which
100 fits are performed for each galaxy.

The results of the application of the algorithm on the LAICA data are
listed in Table~1 and show that both galaxies are quite elliptical,
with axis ratios of 2:1. For Leo~IV, this value is consistent with the
previous measurement based on SDSS data given its large uncertainties,
but the ellipticity--half-light radius covariance \citep[see Fig.~9
of][]{MLfits} implies a system that is significantly larger than
previously measured. As for Leo~V, previous estimates of its
properties relied on a circular model and therefore also
underestimated its size. With the new measurements, both galaxies have
half-light major axes of over 100 pc, removing them from the `size
gap' that exists between dwarf galaxies and globular clusters at
brighter magnitudes \citep[e.g.][]{gilmore07}. The major axes of the
systems do not seem in any way correlated with their relative
positions. The high ellipticity for Leo IV is, however, different from
the results of \cite{sand09}, who find Leo IV to be very round. Since
\cite{sand09} use subgiant and main-sequence stars that are below our
detection threshold and ignore BHB stars, differences could be caused
by morphological differentiations between different stellar
populations. The fact that Leo IV and Leo V lie very close to the
edge, possibly even partly outside of our surveyed area of view might
also influence our measurements in some unforeseen way.

Using the number of stars in each dwarf galaxy, which is an output of
the ML algorithm, and assuming a stellar population (in this case old,
14 Gyr, and metal-poor, [Fe/H]=$-2.3$), the total luminosities are
also derived following the same procedure as in \cite{MLfits}. The
resulting luminosities are $M_V=-5.8\pm0.4$ and $-5.2\pm0.4$ for Leo
IV and V, respectively. The revised value for Leo IV is 0.8 magnitudes
brighter than the value from \cite{MLfits}, but they agree to within
1$\sigma$. It is also statistically equivalent to the value measured
by \cite{sand09} from their deeper data. The difference is due to the
fact that the deeper data reveal that Leo IV is actually more
elongated and has a larger half-light radius than that measured from SDSS
data. Leo V is found to be almost a magnitude brighter than the lower
limit given by \cite{leoV}, who only took into account the light
inside an aperture of 3\arcmin.

\begin{deluxetable}{lcc}
\tablecaption{Structural parameters of Leo IV and Leo V}
\tablewidth{0pt} 
\tablehead{ \colhead{Parameter} & \colhead{Leo IV} & \colhead{Leo V}}
\startdata
$\alpha_0$ (J2000) & 11 32 58.6 $\pm$ 1.6 & 11 31 08.4 $\pm$ 1.6 \\
$\delta_0$ (J2000) & $-$00 33 06 $\pm$ 54 & +02 12 57 $\pm$ 12 \\
$\theta$ (deg) & $-$59 $\pm$ 9 & $-$84 $\pm$ 13 \\
$\epsilon$ & 0.49 $\pm$ 0.11 & 0.50 $\pm$ 0.15 \\
$r_h$ (arcmin) & 4.6$^{+0.8}_{-0.7}$ & 2.6 $\pm$ 0.6 \\
$r_h$ (pc) & 206$^{+36}_{-31}$ & 133 $\pm$ 31 \\
$D$ (kpc) & 154 $\pm$ 5 $^{a)}$ & 175 $\pm$ 9 \\
$m-M$ (mag) & 20.94 $\pm$ 0.07 $^{a)}$ & 21.22 $\pm$ 0.10 \\
$[$Fe/H$]$ & $-$2.3 $\pm$ 0.1 $^{a)}$ & $-$2.0 $\pm$ 0.2 $^{b)}$ \\
$M_V$ & $-$5.8 $\pm$ 0.4 & $-$5.2 $\pm$ 0.4 \\
$L_V$ ($L_\odot$) & $1.8\pm0.8\times10^4$ & $1.0\pm0.5\times10^4$\\
$\mu_V$ (mag/arcsec$^2$) & $27.5\pm0.7$ & $27.1\pm0.8$ \\
$v_\mathrm{GSR}$ ($\kms$) & $10.1\pm1.4^{c)}$ & $60.8\pm3.1^{d)}$ \\
\enddata
\tablecomments{
$^{a)}$\cite{moretti09}; $^{b)}$\cite{walker09};
$^{c)}$\cite{spectro2}; $^{d)}$\cite{leoV}; all other values based
on this publication.
}
\label{tab:structpar}
\end{deluxetable}

\section{A bridge of stars between Leo IV and Leo V?}
\label{sec:stream}

\begin{figure}[t]
\epsscale{1.0}
\plotone{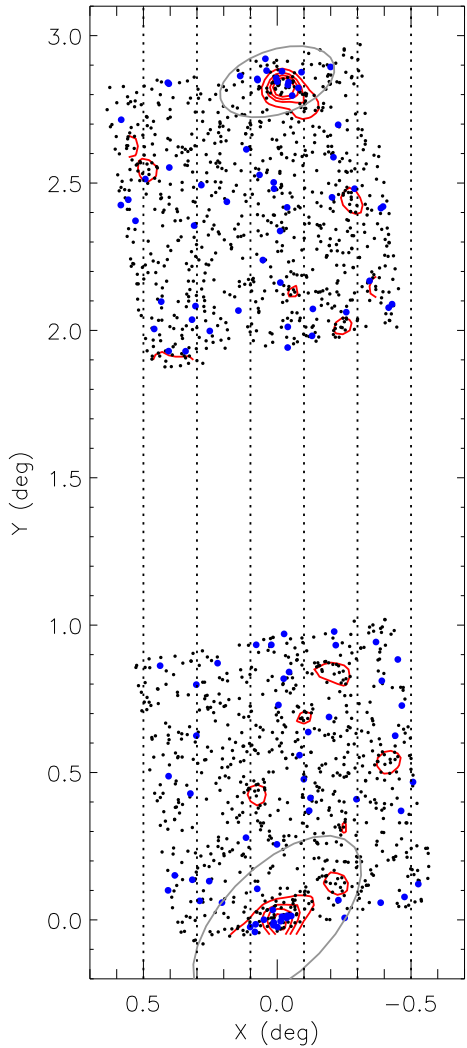} 
\caption{ Spatial distribution of HB (blue dots) and RGB (black dots)
stars around Leo IV and V, selected using the color-magnitude boxes
drawn in Figure \ref{fig:data}. The coordinates are rotated so that
Leo IV and V are both at $X=0$ and Leo IV lies at the origin. Red
contours show the RGB star density, and correspond to
densities exceeding 1.5$\sigma$, 3$\sigma$, 4$\sigma$, and 5$\sigma$
over the background density. The gray ellipses indicate the regions
that are neglected when looking for extra-tidal material, and the
dotted vertical lines outline the bins used for the histograms in
panels (a) to (c) of Fig. \ref{fig:histograms}.
}
\label{fig:targetstars}
\end{figure}

At the Galactic latitudes where Leo IV and V are located,
$b \sim50\deg$, the Galactic stellar populations do not vary
significantly. Hence, in the absence of tidal features related to Leo
IV and/or Leo V the spatial distribution of stars, in any region of
the color-magnitude plane, is expected to be very close to uniform in
the area between the two galaxies. We define a coordinate system
($X,Y$), rotated with respect to ($\alpha$,$\delta$), such that Leo IV
is located at the origin and Leo V lies along the $Y$-axis. Using the
color-magnitude selection boxes overplotted on the CMDs in Figure
\ref{fig:data}, candidate HB and RGB stars at the distance of Leo IV
and V are selected. Their spatial distributions in the ($X$,$Y$)
coordinate system are shown in Figure \ref{fig:targetstars}. The
magnitude limit of $B_0<$23.5 ensures that no systematic completeness
biases affect these distributions (see Fig. \ref{fig:completenessA}
and \ref{fig:completenessC}). Leo IV and Leo V are visible as
overdensities at the very bottom and top of the field. The flattening
of Leo IV is also apparent from the contours showing the RGB star
density, consistent with our structural parameter estimates. However,
no obvious extra-tidal features are visible.

\begin{figure*}[t]
\epsscale{1.0}
\plotone{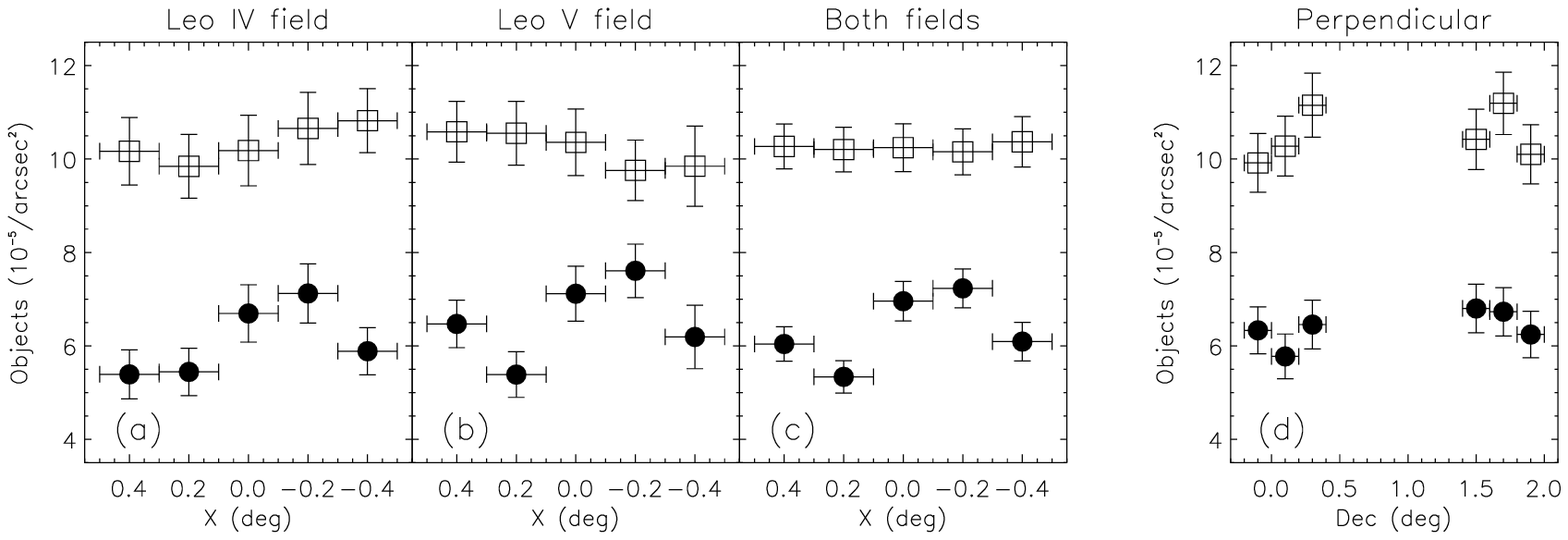}
\caption{ Stellar densities in $0.2\deg$-wide strips. In panels $(a)$,
$(b)$ and $(c)$, the strips are at constant $X$, parallel to the
straight line connecting Leo IV and V. From left to right, the panels
correspond to the field containing Leo IV, the field containing Leo V,
and the sum of both fields. Elliptical regions corresponding to 5
half-light radii around Leo IV and Leo V are avoided
(Fig. \ref{fig:targetstars}). In all panels, filled circles are for
stars selected with the RGB and HB color-magnitude selection boxes
(Fig. \ref{fig:data}) and open squares for foreground M
dwarf stars. Vertical error bars show the Poisson errors and the horizontal
error bars the width of each strip. In panel $(d)$ the strips are at
constant declination $\delta$, almost perpendicular to the straight
line connecting Leo IV and V.  }
\label{fig:histograms}
\end{figure*}

As a sensitive test for a possible overdensity of extra-tidal stars
between the two dwarf galaxies, we divide the observed field in
$0.2\deg$-wide bins along the $X$-axis. In the following analysis we
exclude all data within 5 half-light radii of the dwarf galaxies (the
ellipses in Fig. \ref{fig:targetstars}). The combined density of RGB
and HB stars in each bin is plotted in panels (a) to (c) of Figure
\ref{fig:histograms} with filled circles for the individual mosaics
and the complete data set. The same is done for MW foreground M~dwarf
stars (selected in the color-magnitude region 21.0$<B_0<$23.5,
1.1$<B_0-V_0<$1.6) and plotted with open squares. Similar
$0.2\deg$-wide bins are defined along the $\delta$-axis and their
stellar densities are shown in panel (d). The distribution of MW
foreground stars is consistent with being uniform, both in $X$ and in
$\delta$. In contrast, the RGB and HB stars are uniformly distributed
in $\delta$, but not in $X$, where in both fields substructure is
visible that resembles a bump or overdensity at $X\sim-0.1\deg$. Based
on basic Poisson statistics the significance of this bump is
approximately 3.5$\sigma$. Kolmogorov-Smirnov (KS) tests \citep{numrec}
indicate that the probability of the distribution of RGB and HB stars
along the $X$-axis being drawn from a uniform distribution is only
0.4\%, thus rejecting this null hypothesis at the $\sim3\sigma$
level. Along the $\delta$-axis this probability is 56\%, and for the
MW foreground stars the probabilities are 87\% and 62\% respectively,
thus all consistent with a uniform distribution. Both tests therefore
indicate the presence of substructure only in the distribution of RGB
and HB stars and only along the direction parallel to a line
connecting Leo IV and V. In addition, the KS test also shows that the
distributions of RGB and HB stars along the $X$-axis in the individual
fields are consistent with being drawn from the same distribution,
with a probability of 59\%. The substructure is therefore not isolated
to a single field.

To estimate the significance of this overdensity rigorously, we
performed a ML fit similar to the one used to determine the structural
parameters of the galaxies. The density at any point $(X,Y)$ of the
field of view, $\Sigma(X,Y)$, is here modeled by a flat background,
$\Sigma_b$, with an additional linear enhancement that is constant
along the Leo~IV/V axis (the $Y$-axis of Fig.~\ref{fig:targetstars})
and of Gaussian form $G(X|A,X_0,\sigma)$ along the $X$-axis:
\begin{eqnarray}
\label{eq:model}
\Sigma(X,Y|A,X_0,\sigma) & = & G(X|A,X_0,\sigma) + \Sigma_b\\
 & = & \frac{A}{\sqrt{2\pi}\sigma}\exp\left(-\frac{1}{2}\left(\frac{X-X_0}{\sigma}\right)^2\right) + \Sigma_b.\nonumber
\end{eqnarray}
$A$ denotes the amplitude of the Gaussian, $X_0$ its center
along the $X$-axis and $\sigma$ its standard deviation, also along the
$X$-axis. The probability of the model parameters $(A,X_0,\sigma)$,
given the location $(X_i,Y_i)$ of all considered stars in the data
set, is then
\begin{eqnarray}
p(A,X_0,\sigma|X_i,Y_i, \forall \textrm{ stars } i) \propto \hspace{4cm}\nonumber\\
\mathcal{L}(X_i,Y_i, \forall \textrm{ stars } i|A,X_0,\sigma)\,p(A,X_0,\sigma),\hspace{1.5cm}
\end{eqnarray}
where $\mathcal{L}(X_i,Y_i, \forall \textrm{ stars }
i|A,X_0,\sigma)$ is the likelihood of the model given the data points
and $p(A,X_0,\sigma)$ is the prior probability distribution for the
model's parameters. The likelihood itself is the product, over all
stars, of the model defined in equation (\ref{eq:model}), taken on the
stellar positions. Henceforth
\begin{eqnarray}
\mathcal{L}(X_i,Y_i, \forall \textrm{ stars } i|A,X_0,\sigma)  = \prod_{\mathrm{ star }~i} \Sigma(X_i,Y_i|A,X_0,\sigma)\nonumber\\
= \prod_{\mathrm{ star }~i}\left(\frac{A}{\sqrt{2\pi}\sigma}\exp\left(-\frac{1}{2}\left(\frac{X_i-X_0}{\sigma}\right)^2\right) + \Sigma_b\right).
\end{eqnarray}

The background density, $\Sigma_b$ can also be expressed as a function
of the model parameters by requiring that $N_*$, the total number of
stars in the data set, corresponds to the integral of the density,
$\Sigma$, over the field of view, of area $\mathcal{A}$:
\begin{eqnarray}
\Sigma_b = \Big(N_* - \int_\mathrm{FoV} G(X|A,X_0,\sigma)\,dX dY\Big) \mathcal{A}^{-1}.
\end{eqnarray}

\begin{figure*}[t]
\centering
\includegraphics[angle=0,width=10cm]{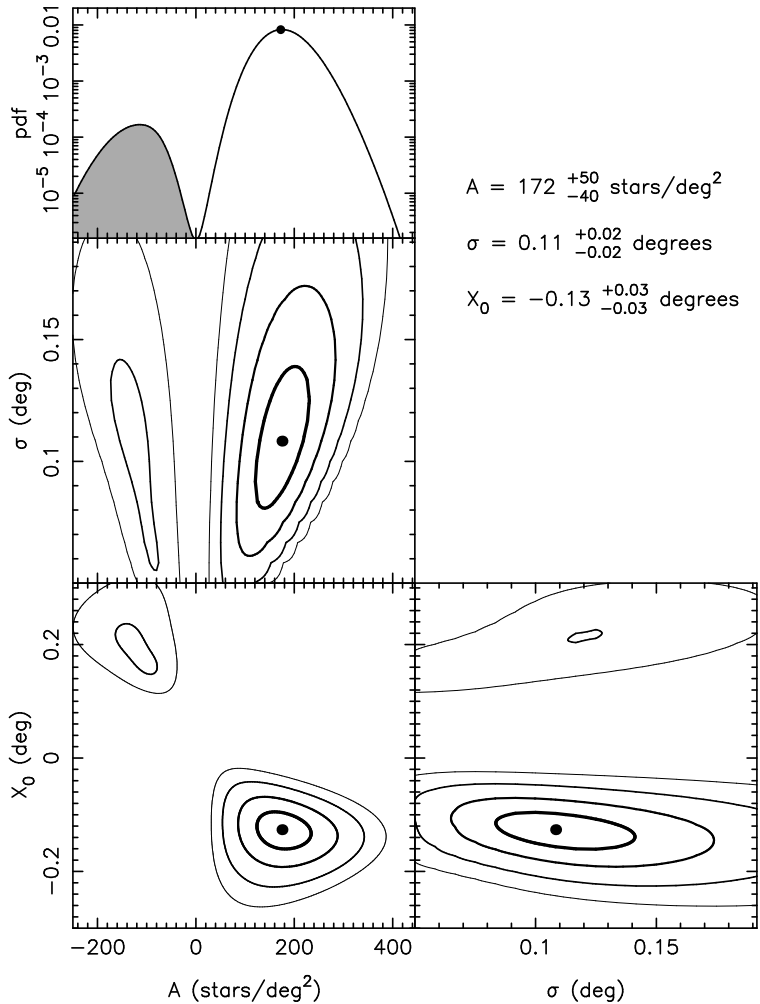} 
\caption{Probability distribution functions from maximum-likelihood
fits of a linear Gaussian to the stellar distribution between Leo IV
and Leo V. {\it Top panel:} probability distribution function of the
model parameter $A$ marginalized over the two other parameters. The
region of negative $A$, shown in grey, yields a small probability of
only 2\%, compared to 98\% for the region of positive $A$; the
probability of the best model (indicated with a filled circle in all
panels) is a factor $\sim$10$^{4}$ higher than that of a uniform
distribution ($A$=0), which corresponds to a $\sim$4$\sigma$
significance.  {\it Bottom panels:} for all combinations of the three
fit parameters ($A$, $X_0$ and $\sigma$) the two-dimensional
probability function is shown, marginalized over the third
parameter. Here, contours represent drops of 50\%, 90\%, 99\% and
99.9\% of the probability with respect to the best model.}
\label{fig:MLresults}
\end{figure*}

The ML fits benefit from a larger sample of stars than provided by the
selection boxes in Figure \ref{fig:data} to constrain the fit
parameters optimally. Here we therefore use a different selection box
(the dashed box in Figure \ref{fig:data}) containing 5\,891~stars in
the surveyed area but more than $5r_h$ from the center either of the
two dwarf galaxies.  The prior probability distributions for the
model's parameters were chosen to be uniform over ranges that are
physically motivated by a stream of stars originating from the dwarf
galaxies: the probability on $\sigma$ is uniformly distributed between
$3'$ (the typical size of the dwarf galaxies) and $12'$, whereas $X_0$
and $A$ are uniformly chosen between $\pm0.5\deg$ (the width of the
data set), and $\pm450$, respectively. From these, we find that the
most probable model has the following properties: $X_0=-0.13 \pm
0.03\deg$, $\sigma = 0.11 \pm 0.02\deg$, and
$A=172^{+50}_{-40}$\,stars/deg$^2$. This corresponds to a background
density of $\Sigma_b=3224$\,stars/deg$^2$ for the chosen selection box
and a total of 275 stars in the overdensity for the selection box and
footprint used. Comparing the peak amplitude of the overdensity to the
number of RGB and BHB stars in Leo IV yields a central surface
brightness of $\mu_V\simeq$32 mag arcsec$^{-2}$.  The 2-D probability
distribution functions are shown in Figure~\ref{fig:MLresults} for all
parameters. The significance of the peak in the probability
distribution function is just over 4$\sigma$ in all cases, leading us
to conclude that the data show evidence of a 4$\sigma$ overdensity of
stars that coincides with the bump visible in
Figure~\ref{fig:histograms}.

\section{Discussion}
\label{sec:discussion}

All tests --- the histograms in Figure \ref{fig:histograms}, the KS
tests and the ML fits --- indicate that there is structure in the distribution
of stars consistent with being Leo IV or Leo V members in the area 
between the systems. The strongest case is provided by the last test,
which indicates a significance of the overdensity of
$\gtrsim$4$\sigma$.  Aside from the preferred model, there is another
local maximum visible in Figure \ref{fig:MLresults} that corresponds
to an underdensity in the data set. This coincides with the drop in
density visible at $X_0\simeq+0.2\deg$ in panels (b) and (c) of
Figure~\ref{fig:histograms}. The limited extent of our survey
perpendicular to the possible bridge of stars implies that an
overdensity can also be fit as a region with high background, with an
adjacent underdensity. However, marginalizing over all parameters but
$A$ (top panel of Fig.~\ref{fig:MLresults}) makes it clear that the
probability of this `underdensity' is only 2\%, whereas the
probability of an overdensity corresponds to 98\%. This shows that the
structure in the field is more likely caused by an overdensity than by
an underdensity. This inference is based purely on statistics, not
taking into account that it is not clear what could cause an
underdensity of stars in the field. 

In their recent analysis, \cite{sand09} find no evidence for
extra-tidal features emanating from Leo IV. Due to their much smaller
field-of-view their analysis is, however, considerably less sensitive
to the presence of a possible tidal stream, despite deeper photometry.
In the region between Leo IV and Leo V, where any putative bridge
would be found, our survey covers an area seven times larger than that
available to \cite{sand09}. At a given depth, this leads to a
seven-fold increase in the number of stellar tracers, in our case RGB
and BHB stars, that can be found. The higher sensitivity of the
\cite{sand09} data only partly compensates for this, and only adds
fainter stars in a region of the CMD where contamination by faint
galaxies is high. The detection limit of \cite{sand09} for tidal
streams of $\mu_g \lesssim$29.8 mag arcsec$^{-2}$ is insufficient to
detect the stellar bridge, which has an estimated central surface
brightness of $\mu_V \sim$32 mag arcsec$^{-2}$. It is, however,
interesting to note that one of the significant `nuggets' of stars
seen by \cite{sand09} to the north-west of Leo IV lies exactly on top
of the location of the stellar bridge, as indicated by our ML fits.

The spatial proximity of Leo IV and Leo V and the intriguing
possibility of extra-tidal material between them elicits the question
of whether these two systems are somehow related and/or interacting
with each other. Here we consider the following different scenarios:
\begin{itemize}
\item Leo IV and Leo V might be condensations of stars in a much
larger stellar stream, even though the density contrast between the
condensations and the underlying stream would be very high. The
apparent width of the extra-tidal substructure in the candidate RGB
and HB stars of $\sim$15 arcmin is similar to the sizes of Leo IV and
V, and is therefore consistent with this scenario. In this case they
would be expected to closely follow the same orbit, which should come
close enough to the Galactic center to cause the tidal disruption of a
common progenitor.
\item Leo IV and Leo V might have fallen into the Milky Way together
as members of a group of dwarf galaxies, a process expected to be
common in a $\Lambda$CDM cosmology \citep[e.g.][]{donghia08,li08}, or
as satellites of a larger halo, as has been suggested for Segue 1 and
2 \citep{segue2}. Regarding the latter case, it should be noted that
the revised half-light radii of Leo IV ($\sim$200 pc) and Leo V
($\sim$130 pc) make them much larger systems than Segue 1 and 2 and,
in terms of size, clearly more similar to other dSphs. If they do
constitute a pair with a common origin, they would still follow the
same orbit around the Milky Way, but with a possibly large velocity
dispersion, as they orbit a common center of mass. In this scenario,
extra-tidal stars might be the result of interactions between the two
galaxies, rather than with the Milky Way.
\item A final option would be that Leo IV and Leo V were originally
unrelated satellites of the Milky Way, but had a chance encounter. A
near head-on collision might be able to cause an interaction strong
enough to disrupt their stellar bodies.
\end{itemize}

\subsection{A common orbit}

If Leo IV and Leo V are tied together and follow the same orbit around
the Milky Way, their common orbit should be easily determinable from
simple physical arguments. In the following, we assume that the
Galactocentric properties of the galaxies are interchangeable with
their heliocentric properties, a reasonable assumption given the large
distances to the systems compared to the solar radius around the Milky
Way. If the two dwarf galaxies lie along the same orbit, they should
actually share the same orbital energy and angular momentum. The
latter requirement can first be used to link together the tangential
velocities of the two galaxies in the plane of the sky, $v_{t,4}$ and
$v_{t,5}$, and their distances, $r_4$ and $r_5$, respectively for
Leo~IV and Leo~V: $v_{t,5} = v_{t,4} r_4/r_5$. We then invoke energy
conservation along the Leo~IV/V orbit to obtain
\begin{equation}
\left(v_{\mathrm{tot},4}\right)^2 - \left(v_{\mathrm{esc},4}\right)^2 = \left(v_{\mathrm{tot},5}\right)^2 - \left(v_{\mathrm{esc},5}\right)^2,
\end{equation}
where $v_{\mathrm{tot},i} = \sqrt{(v_{t,i})^2 + (v_{r,i})^2}$ is the
total velocity of Leo~$i$, $v_{r,i}$ its radial velocity with respect
to the Galactic Standard of Rest, and $v_{\mathrm{esc},i}$ the escape
velocity in the assumed model of the Milky Way potential, at the
location of Leo~$i$. For Leo~IV, the tangential motion of the orbit
that links Leo~IV and Leo~V is therefore defined by:
\begin{eqnarray}
\label{eq_vt}
v_{t,4} =\hspace{8cm}\nonumber\\
\sqrt{\left(\left(v_{r,5}\right)^2 -
  \left(v_{r,4}\right)^2+\left(v_{\mathrm{esc},4}\right)^2-\left(v_{\mathrm{esc},5}\right)^2\right) \left(1-\left(\frac{r_4}{r_5}\right)^2\right)^{-1}}.
\end{eqnarray}

In order to determine $v_{t,4}$ and $v_{t,5}$ with their associated
uncertainties, we follow a Monte Carlo scheme and randomly generate
the distances and radial velocities for each of the two dwarf galaxies
from Gaussian distributions defined by their observed properties, as
listed in Table \ref{tab:structpar}. We then determine the escape
velocities required by equation (\ref{eq_vt}) for the Milky Way model
adopted by \citet{paczynski90}, after removing unphysical cases where
one of the two galaxies is not bound to the Milky Way. The median and
central 68.3\% of the resulting distributions after 10\,000 trials
yield $v_{t,4} = 239_{-6}^{+19}\kms$ and $v_{t,5} =
211_{-15}^{+30}\kms$.

\begin{figure}[t]
\epsscale{0.9}
\plotone{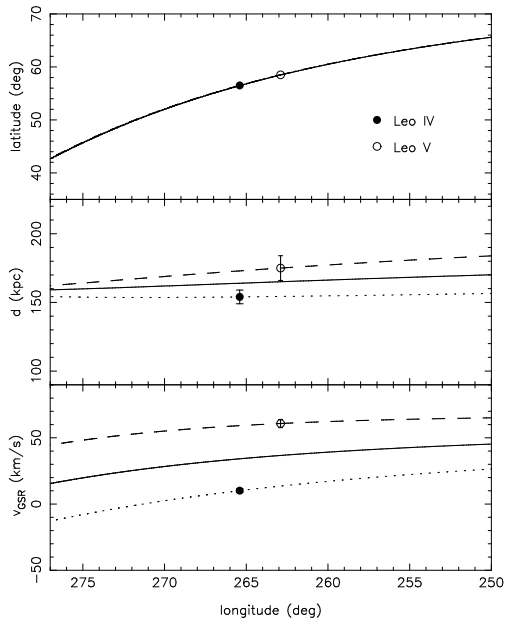}
\caption{ Orbits for Leo IV and V, derived assuming common energy and
angular momentum for the two dwarf galaxies. The dotted, dashed and solid
lines correspond to orbits with initial conditions determined using
Leo~IV, Leo~V or their mean properties, respectively. The panels show
the evolution of Galactic latitude (top), heliocentric distance
(middle) and radial velocity with respect to the Galactic Standard of
Rest (bottom) as functions of Galactic longitude. Properties of Leo~IV
and V are indicated by filled and open circles, respectively. }
\label{fig:orbits}
\end{figure}

These large values, combined with the small radial velocities of the
dwarf galaxies, make it impossible to find a viable orbit that
reproduces the observed properties of both systems at the same time.
None of the orbits drawn in Figure~\ref{fig:orbits} goes through both
points in all panels, demonstrating that, whether Leo~IV or Leo~V is
used as a starting point, neither the velocity nor the distance along
the orbit has a strong enough gradient to replicate the observed
properties of the other dwarf galaxy. This analysis thus shows that
Leo~IV and~V cannot be following the exact same orbit. We can
calculate a compromise solution, with a starting point for the orbit
integration that has the mean properties of the two galaxies (the
central line in Figure~\ref{fig:orbits}). This orbit does not reproduce
the velocities and distances of the two systems and remains at large
Galactocentric distances, with an apocenter of $244\kpc$ and a
pericenter of $158\kpc$. We conclude that Leo IV and Leo V
are very unlikely to be condensations in a larger stellar stream, for
two reasons: no viable single orbit exists, and the compromise
solution does not come close enough to the Galactic center for a
single progenitor to have been tidally affected.

\subsection{A `tumbling pair'}

Rather than being remnants of a progenitor disrupted by the
gravitational tides of the Milky Way, the two dwarf galaxies could be
a gravitationally bound pair interacting with each other. This means
that they would follow a `common' orbit, but with a large velocity
dispersion, as they orbit their common center of mass. Such a
`tumbling pair' of dwarf galaxies would be similar to the Large and
Small Magellanic Clouds, albeit much more scaled-down. For the two
systems to be gravitationally bound, the total kinematic energy of the
system must be less than the total gravitational potential energy. As
pointed out by \cite{davis95}, the two-body Newtonian binding criterion
can be written as
\begin{equation}
M_{sys} \geq \frac{R v_{los}^2}{2 G \sin^2 \alpha},
\end{equation}
where $M_{sys}$ is the total mass of the system, $R$ the distance
between the two objects, $v_{los}$ the line-of-sight relative
velocity, and $\alpha$ the angle between the axis of the two-body
system and the sky. Inserting the values for these parameters
($v_{los}=50.7 \pm 3.4$ km s$^{-1}$, $R=22 \pm 10$ kpc,
$\alpha=70$\degr$^{+6}_{-18}$; see Table \ref{tab:structpar} and
references therein) yields a lower limit for the total mass of
$M_{sys} \geq 8 \pm 4 \times 10^9$ M$_\sun$, or half of that for each
individual dwarf.

Masses inferred from the stellar velocity dispersions in faint dSph
galaxies are typically much lower, but only probe the inner few
hundred parsec of their dark matter halos. Based on dynamical modeling
of dSphs embedded in cosmologically motivated dark matter halos,
\cite{penarrubia08} derive masses of up to several times $10^9$ M$_\sun$
for Local Group dSphs. The masses required for Leo IV and Leo V to be
a bound pair are therefore possible, but imply high mass-to-light
ratios of M/L$_V$=10$^5$ in solar units.

\subsection{A close encounter}

\begin{figure}[t]
\epsscale{1.0}
\plotone{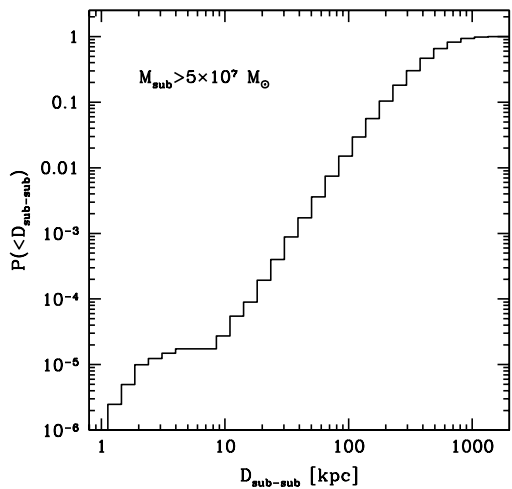}
\caption{Cumulative probability distribution for the distance between
two dark matter subhalos (with a current mass larger than $5 \times
10^7 M_{\odot}$) in the last 3.4 Gyr (see text for details).  }
\label{fig:maccio}
\end{figure}

A third option would be that Leo IV and V are unrelated systems that
randomly collided in the Milky Way halo, thus provoking an
interaction. Given that the objects have a stellar extent of $\sim$500
pc, we estimate that they would need to come to within a few
kiloparsecs of each other with a relative velocity of less than
$\sim100 \kms$ for tides to affect their stellar components. We used
results from N-body simulations in order to compute the probability of
such an encounter. We started from the four Milky Way-like dark matter
($\Lambda$CDM) halos presented in \cite{maccio09} and, for each
subhalo within 250 kpc of the host galaxy's center and with a bound
mass today $M_{sub} > 5 \times 10^7 M_{\sun}$, we determined the
distance to any other subhalo, $D_{sub-sub}$, in the last 3.4 Gyr.
Figure \ref{fig:maccio} shows the cumulative probability distribution
for $D_{sub-sub}$ in the four Milky Way analogue simulations.  The
probability of two halos coming to within 5 kpc of one another is as
low as a few times $10^{-5}$. In addition, this number can only be an
upper limit for a galaxy-galaxy interaction, since only $\approx 30\%$
of dark matter halos with $M_{sub} > 5 \times 10^7 M_{\sun}$ are
expected to host stars \citep{maccio09}.  We conclude that a collision
between Leo IV and Leo V is not a likely explanation for a possible
interaction.

\section{Conclusions}
\label{sec:conclusions}

Given that Leo IV and Leo V form a close pair on the sky, in distance,
and in radial velocity, it has been argued that they might be
interacting or even share the same orbit \citep{leoV}.  We have
obtained new, deep photometry in two $1\deg\times1\deg$ fields
containing both dwarf spheroidal galaxies, re-derived their structural
properties, and searched for extra-tidal stars. Both Leo IV and Leo V
are considerably larger than previously thought, with sizes of 206 pc
and 133 pc. These sizes are similar to those of other recently
discovered dSphs such as Bo\"otes I or Ursa Major II
\citep[respectively 242 and 140 pc,][]{MLfits} and the smaller
`classical' dSphs, for example Draco \citep[221 pc,][]{MLfits} or
Leo II \citep[185 pc,][]{coleman07}. Both are also strongly elongated,
with axis ratios of 2:1, which is the case for many of the fainter
($M_V>-8$) dSphs \citep{MLfits}. We note, however, that the
proximity of both dwarf galaxies near the edge of our survey might
have some influence on these measurements.

The combination of deep photometry and a large area coverage of our
survey provides an unprecedented sensitivity to low surface brightness
features in the region between Leo IV and Leo V.
Analysis of the spatial distribution of candidate RGB
and HB stars using density histograms and Kolmogorov-Smirnov tests
reveals the presence of a significant substructure in the observed
fields. This substructure is detected at the $\sim3\sigma$ level, but
only along the direction connecting the two systems. Maximum-likelihood
modeling of the stellar distribution with a linear Gaussian confirms
that the spatial structure can be fit with an overdensity connecting
the two systems. With this method, the significance of the overdensity
is just over 4$\sigma$, and has a peak surface brightness of
$\mu_V\simeq$~32 mag arcsec$^{-2}$.

The reason for the close proximity of Leo IV and Leo V is still an
enigma, and we have considered a few possible explanations.  Orbits
derived for the objects using their positions and velocities and by
requiring the pair to share common values for energy and angular
momentum show that it is not possible for both to be on the exact same
orbit under reasonable assumptions. Furthermore, a compromise orbit
would not approach the Galactic center closer than $\sim$160 kpc, as
also noted by \cite{leoV}. Hence, it is highly unlikely that Leo IV
and Leo V are condensations in a low surface brightness stellar stream
resulting from the tidal disruption of a common progenitor. The fact
that the elongations of the two systems are not aligned might be a
further indication of this. A large velocity dispersion around a
mutual orbit, for example in the case of a `tumbling pair' of dwarf
galaxies, is only viable if Leo IV and Leo V have masses of at least a
few times 10$^9$ M$_\sun$. This would imply extreme $V$-band
mass-to-light ratios of $\gtrsim$10$^5$ in solar units, but recent
dynamical simulations of dwarf spheroidal galaxies
\citep{penarrubia08} show that the required masses are possible. If
Leo IV and Leo V are indeed a bound and internally interacting pair,
this could account naturally for their unaligned elongations and the
offset of the apparent `stellar bridge' from the straight line
connecting them. Finally, analysis of Milky Way-like $\Lambda$CDM
halos reveals that the probability of two satellites colliding is
negligibly small. This is consistent with the findings of \cite{leoV}
who calculate a $\lesssim$1\% probability for such a close association
happening by chance. Therefore, the scenario in which Leo IV and Leo V
are a bound pair of dwarf galaxies, orbiting and interacting with each
other, appears to be the most viable explanation for this close
celestial companionship.

\acknowledgments
The authors thank the anonymous referee for helpful comments.
Based on observations collected at the Centro Astron\'omico Hispano
Alem\'an (CAHA) at Calar Alto, operated jointly by the Max-Planck
Institut f\"ur Astronomie and the Instituto de Astrof\'isica de
Andaluc\'ia (CSIC). This research has made use of the VizieR catalogue
access tool, CDS, Strasbourg, France.

\end{document}